\documentclass[letter]{ieice}
\pdfoutput=1
\usepackage[]{graphicx,xcolor}
\usepackage[fleqn]{amsmath}
\usepackage{newtxtext}
\usepackage[varg]{newtxmath}
\usepackage[]{caption}
\usepackage[]{subcaption}
\usepackage{multirow}
\field{}

\title{A Prompt Report on the Performance of Intel Optane DC Persistent Memory Module}
\authorlist{%
 \authorentry{Takahiro Hirofuchi}{n}{aist}\MembershipNumber{}
 \authorentry{Ryousei Takano}{m}{aist}\MembershipNumber{1617664}
}
\affiliate[aist]{National Institute of Advanced Industrial Science and Technology (AIST)}

\begin{document}
\maketitle
\begin{summary}
In this prompt report, we present the basic performance evaluation of
Intel Optane Data Center Persistent Memory Module (Optane DCPMM), which
is the first commercially-available, byte-addressable non-volatile
memory modules released in April 2019. Since at the moment of writing
only a few reports on its performance were published, this letter is
intended to complement other performance studies.
Through experiments using our own measurement tools, we
obtained that the latency of random read-only access was approximately 374 ns.
That of random writeback-involving access was 391 ns. The bandwidths of
read-only and writeback-involving access for interleaved memory modules
were approximately 38 GB/s and 3 GB/s, respectively.
\end{summary}

\begin{keywords}
Non-volatile Memory, NVM, Optane DC PM, DCPMM
\end{keywords}

\section{Introduction}
In April 2019, Intel officially released the first commercially-available,
byte-addressable NVM technology, Intel Optane Data Center Persistent
Memory Module (DCPMM). DCPMM is a long-awaited product drastically increasing main memory capacities.
Since DRAM technology is unlikely able to meet this growing memory demand,
non-volatile memory (NVM) technologies, being accessible in the same manner as DRAM,
are considered indispensable for expanding main memory capacities.
However, there is a substantial performance gap between DRAM and DCPMM.

Since DCPMM was released, only a few reports on its performance were published (\cite{dcpmmucsd1903,dcpmmmut1904}).
This prompt report is intended to complement other performance reports on DCPMM
and pave the way for further system software studies addressing the performance
gap.
We developed our own micro-benchmark programs to measure memory latency and bandwidth
and investigated bare performance of DCPMM to see its fundamental characteristics.
~\footnote{Note that Intel Optane DCPMM (released in 2019) and Intel Optane Memory (released in 2017) are different products. The latter is a storage class memory device connected to the PCIe NVMe interface. DCPMM is connected to the DIMM interface and seen as main memory from CPU if configured as the App Direct mode.}
\footnote{To promptly report results and obtain feedback from the community, we uploaded the early summary of our experiments to a public preprint server \cite{dcpmmaist1907}. It summarizes the basic performance of DCPMM as well as its feasibility to our hypervisor-based virtualization mechanism for hybrid memory systems. Considering the broader reader's interest and the page limit of the IEICE letter format, we focus this paper only to the results of basic performance evaluation. In this letter, we added discussion on how this work complements other performance reports.}

To clarify the contribution of this letter, we summarize our obtained performance numbers and compare them with the ones reported by related work:
\begin{itemize}
\item Although \cite{dcpmmucsd1903} reported that the read latency of DCPMM is 305 ns,
we obtained 374 ns, which is close to 391 ns reported by \cite{dcpmmmut1904}. As discussed later, there is a possibility that the measurement tool used in \cite{dcpmmucsd1903} (i.e., Intel MLC v3.6) outputted a relatively small value.
\item In \cite{dcpmmucsd1903} and \cite{dcpmmmut1904}, the write latency of DCPMM was measured with non-temporal instructions or cache-control instructions (e.g., clflush).
Although depending on conditions, their values were generally in the range of 100-200 ns.
On the other hand, we conducted experiments from another viewpoint, in order to see write latencies possibly experienced by ordinary applications (that do not intentionally use non-temporal and cache-control instructions for NVM).
The estimate value of its write latency through our experiments was 391 ns.
Considering the write mechanism of the 3D Xpoint technology, it is very unlikely that its actual write latency is much shorter than its read latency.
Possibly,
the write latencies obtained by non-temporal and cache-control instructions present a period of time to deliver data to the non-volatile internal buffer of a memory controller or memory module (that ensures no data loss upon a power failure), which is not a period of time to actually deliver data to non-volatile memory cells.

\item Regarding the read bandwidth of DCPMM, \cite{dcpmmucsd1903} reported 39.4 GB/s by measuring the performance of sequential read with Intel MLC v3.6. \cite{dcpmmmut1904} reported 37 GB/s by measuring random read at the granularity of 4 adjacent cache lines. We obtained 37.6 GB/s by doing experiments in which multiple worker processes performed sequential read on each non-overlapped scratch buffer. Our result corroborates the already reported performance numbers.

\item Regarding its write bandwidth, \cite{dcpmmucsd1903} reported 13.9 GB/s by Intel MLC v3.6. \cite{dcpmmmut1904} reported 4 GB/s. In our experiments, the peak performance was 3 GB/s. Although the details of the measurement algorithm of Intel MLC were not available, we consider that 13.9 GB/s was an unlikely high value, which will not represent time to actually reach memory cells.
Our result was more conservative than \cite{dcpmmmut1904}.

\item While interleaving was not disabled in \cite{dcpmmucsd1903} and \cite{dcpmmmut1904}, we also measured the read/write bandwidths and latencies with non-interleaved configurations. For example, the read/write latencies were degraded by 5.4\% and 17.2\%, respectively. Since interleaving contributed to decreasing latencies, there will be multiple request queues to access memory modules.
As the number of concurrent reading processes increased, the read bandwidth drastically decreased.
We observed this behavior only in the case of read access with interleaving disabled.
Although it is difficult to explain the exact reason of this behavior because the technical detail of DCPMM is not disclosed, a possible reason is that its internal buffering mechanism does not work efficiently when the interleaving mechanism is disabled.

\end{itemize}

\begin{table}[t]
	\begin{center}
	\caption{The overview of the test machine used in experiments}
	\label{tbl:machine}
	{\footnotesize
	\begin{tabular}{l l}
		\hline
		CPU & Intel Xeon Platinum 8260L 2.40 GHz (Cascade Lake) x2\\
		    & L1d cache 32 KB, L1i cache 32 KB \\
		    & L2 cache 1024K \\
		    & L3 cache 36 MB \\
		DRAM & DDR4 DRAM 16 GB, 2666 MT/s, 12 slots \\
		DCPMM & DDR-T 128 GB, 2666 MT/s, 12 slots \\
		OS & Linux Kernel 4.19.16 (extended for RAMinate) \\
		\hline
	\end{tabular}
	}
		\end{center}
\end{table}

\begin{figure}[t]
	\begin{center}
	\includegraphics[width=0.6\columnwidth]{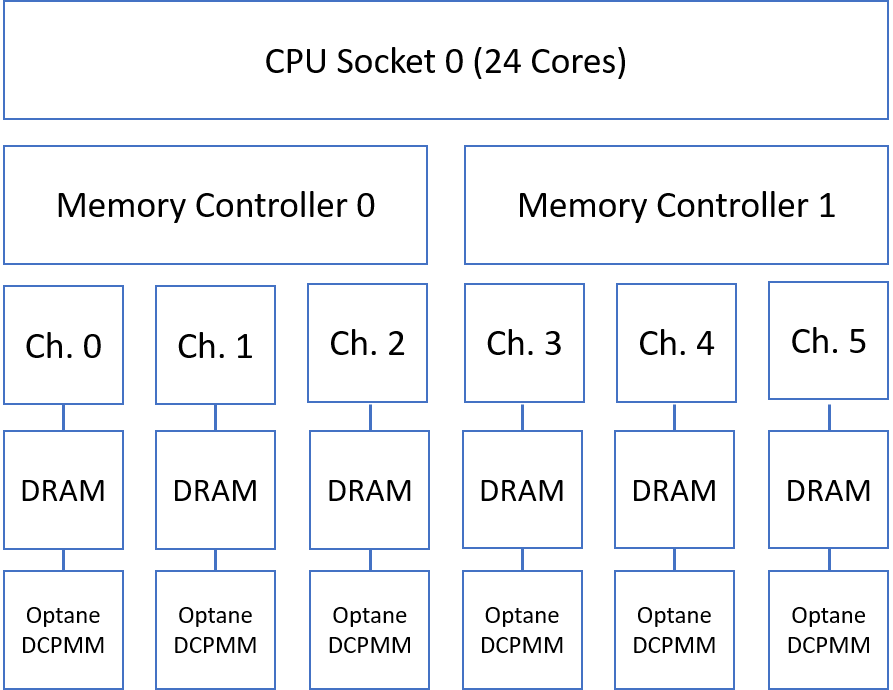}
	\caption{The memory configuration of the tested machine (NUMA 0)}
	\label{fig:memtopo}
	\end{center}
\end{figure}

\section{Evaluation}
\label{sec:eval}

Table \ref{tbl:machine} summarizes the specification of the tested machine.
Figure \ref{fig:memtopo} shows its memory configuration.
The machine is equipped with 2 CPU sockets. A CPU processor has 24 physical CPU cores and 2 memory controllers.
A memory controller has 3 memory channels. Each memory channel has a DDR4 DRAM module (16 GB) and a DCPMM (128 GB).
The total DRAM size of the machine is 192 GB. The total DCPMM size is 1536 GB.

The Intel CPU processors supporting DCPMM allow users to configure how DCPMM is
incorporated into the main memory of a computer. In experiments,
we assigned all the DCPMMs to App Direct Mode.
In App Direct Mode, the memory controller maps both DRAM and DCPMM to the
physical memory address space of the machine, which enables the software layer to directly accesses DCPMM.

The host operating system leaves DCPMMs intact.
The benchmark programs directly accessed the physical memory ranges of DCPMMs via the
device file of Linux ({\tt /dev/mem}).
Although the operating system recognized two NUMA domains (i.e., those of CPU
socket 0 and 1, respectively), we used the CPU cores and memory modules only in
the first NUMA domain.

The interleaving mechanism of DRAM and that of DCPMM were enabled, respectively.
For DCPMM, the interleaving configuration of App Direct Mode was used unless otherwise noted.
The 6 DCPMMs connected to each NUMA domain were logically combined.
The memory controller spread memory accesses evenly to the memory modules.
For DRAM, the controller interleaving (i.e., iMC interleaving) was enabled in the BIOS setting.
Similarly, the 6 DRAM modules connected to each NUMA domain were logically combined.
In order to simplify system behavior, we disabled the hyper-threading mechanism of CPUs.
Transparent huge page and address randomization were also disabled in the setting of Linux Kernel.

We developed micro-benchmark programs that measure the read/write access
latencies and bandwidth of physical memory\footnote{The micro-benchmark programs were also used in our prior studies. Refer to \cite{mesmeric2019} for more information.}.
To measure read performance, the micro-benchmark programs induce Last Level Cache (LLC) misses that result in data fetches from memory modules.
For write performance, the programs cause the evictions of modified cachelines as well.

\begin{figure*}[t]
	\begin{center}
	\includegraphics[width=1.8\columnwidth]{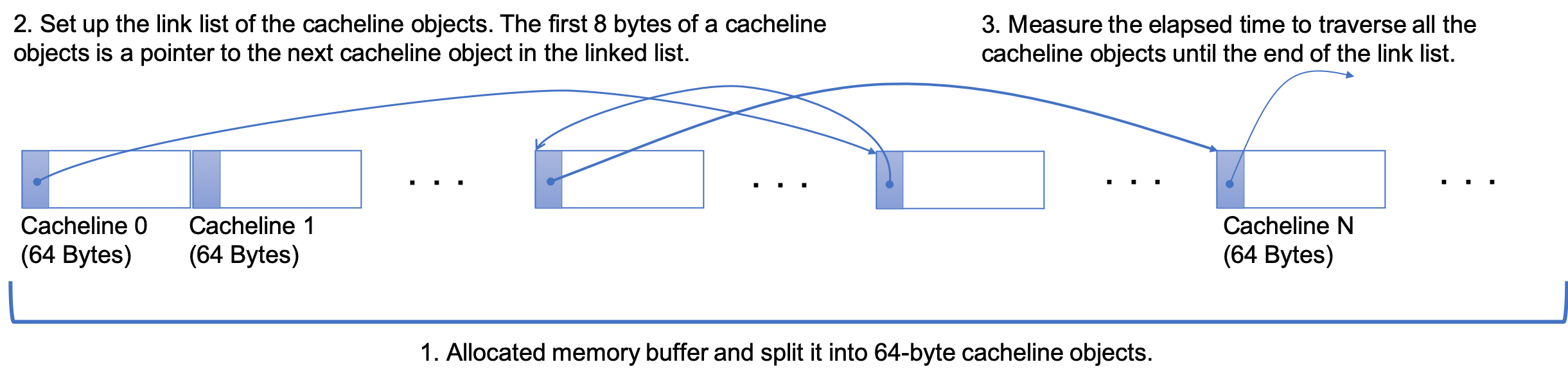}
	\caption{The overview of the micro-benchmark program to measure memory read/write latencies}
	\label{fig:wbbench}
	\end{center}
\end{figure*}

\subsection{Read/Write Latencies}
\label{sec:explat}

Figure \ref{fig:wbbench} illustrates the overview of the micro-benchmark
program to measure memory read/write latencies.
Most CPU architectures perform the memory prefetching and the out-of-order execution to
hide memory latencies from programs running on CPU cores.
To measure latencies precisely, the benchmark program was carefully designed to suppress these effects.
To measure the read latency of main memory, it works as follows:
\begin{itemize}
	\item First, it allocates a certain amount of memory buffer from a target memory device. To induce LLC misses, the size of the allocated buffer should be sufficiently larger than the size of LLC. It splits the memory buffer into 64-bytes cacheline objects.
	\item Second, it set up the link list of the cacheline objects in a random order, i.e., traversing the linked list causes jumps to remote cacheline objects.
	\item Third, it measures the elapsed time for traversing all cacheline objects and calculates the average latency to fetch a cacheline.
		In most cases, a CPU core stalls due to an LLC miss upon the traversal of the next cacheline object in the linked list. The elapsed time of this CPU stall is a memory latency.
\end{itemize}

When measuring the write-back latency, in addition to the second step, it updates the second 8 bytes of a cacheline object before jumping to the next cacheline object.
The status of the cacheline in LLC changes to {\it modified}. The cacheline is written back to main memory later.
Although a write-back operation is asynchronously performed,
we can estimate the average latency of a memory access involving the write-back of a cacheline, from the elapsed time to traverse all the cache link objects.

\begin{figure*}[t]
	\begin{center}
	\begin{subfigure}{0.33\linewidth}
		\includegraphics[width=\columnwidth]{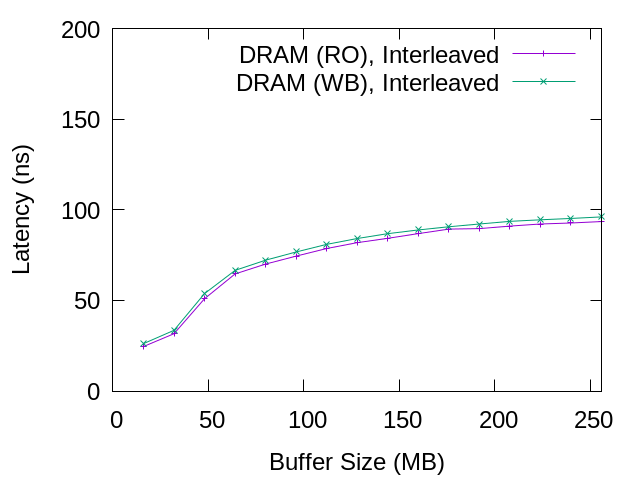}
		\caption{DRAM (Interleaved)}
	\end{subfigure}
	\begin{subfigure}{0.33\linewidth}
		\includegraphics[width=\columnwidth]{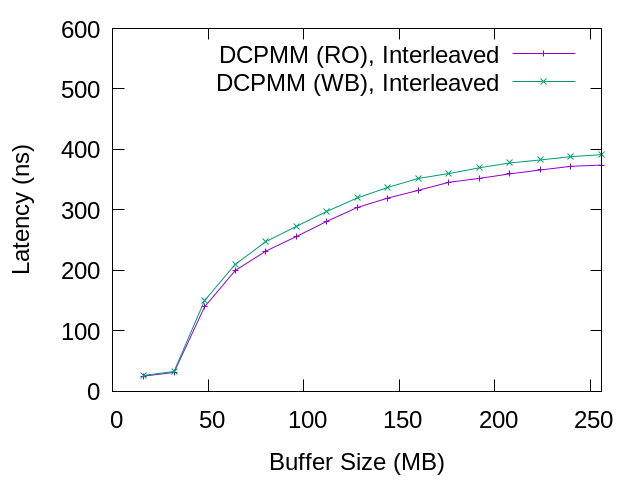}
		\caption{DCPMM (Interleaved)}
	\end{subfigure}
	\begin{subfigure}{0.33\linewidth}
		\includegraphics[width=\columnwidth]{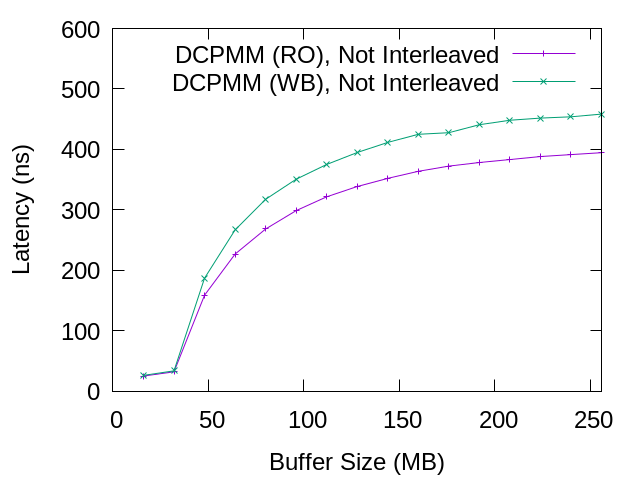}
		\caption{DCPMM (Non Interleaved)}
	\end{subfigure}
	\caption{The read and write latencies of DRAM and DCPMM.
	In the graphs, the results of the read latency are marked as RO
	(read-only), and those of the write latency are marked as WB
	(write-back).}
	\label{fig:memlat}
	\end{center}
\end{figure*}

Figure \ref{fig:memlat} summarizes the measured results of the read/write latencies of DRAM and DCPMM, respectively.
As the size of the allocated memory buffer increased,
the read/write latencies of DRAM reached approximately 95 ns, respectively.
Although write latencies were slightly higher with any tested buffer sizes,
the differences in read/write latencies were only 1-2 ns.
On the other hand, the read latency of DCPMM was up to 374.1 ns. The write latency was 391.2 ns.
For read access, the latency of DCPMM was 400.1\% higher than that of DRAM. For write access, it was 407.1\% higher.
Similarly to other NVM technologies, the write latency of a bare DCPMM module was larger than the read latency, as clearly shown in the result of the non-interleaved configuration. The latency of memory access involving write-back was 458.4 ns, which was 16.1\% higher than that of read-only access (394.5 ns).
The read/write latency was degraded by 5.4\% and 17.2\%, respectively, in comparison to the interleaved cases.

It should be noted that these measured latencies include the penalty caused by
TLB (Translation Lookaside Buffer) misses. The page size in the experiments was 4 KB.
Our measured latencies of DRAM were slightly higher than the value that Intel Memory Latency Checker (MLC) reported.
Intel MLC v3.6 reported that the DRAM latency was 82 ns.
The method of random access in Intel MLC slightly differs from that of our
micro-benchmark program. According to the documentation of Intel MLC v3.6, it performs
random access in a 256-KB range of memory in order to mitigate TLB misses. After
completing that range, it performs random access in the next 256-KB range of
memory.
We consider that memory intensive applications randomly accessing a wide range
of memory will experience memory latencies close to our obtained results.
Although it is out of the scope of this report,
one could use a large page size such as 2 MB and 1 GB to mitigate TLB misses.

\subsection{Read/Write Bandwidths}
Our micro-benchmark program measuring the read/write bandwidths of main memory
launches a multiple number of concurrent worker processes to perform memory access.
Each worker process allocates 1 GB of memory buffer from a target memory device.
The memory buffer of a worker process does not overlap the memory buffer of another worker process.
Each worker process sequentially scans its allocated buffer.
We increased the number of worker processes up to the number of CPU cores of an NUMA domain.

\begin{figure*}[t]
	\begin{center}
	\begin{subfigure}{0.33\linewidth}
		\includegraphics[width=\columnwidth]{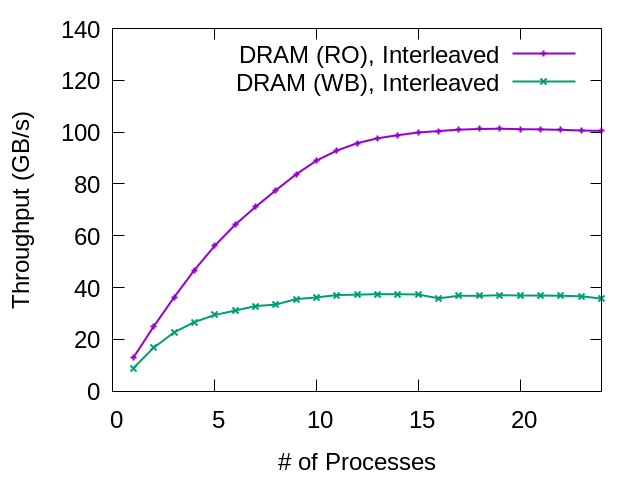}
		\caption{DRAM (Interleaved)}
	\end{subfigure}
	\begin{subfigure}{0.33\linewidth}
		\includegraphics[width=\columnwidth]{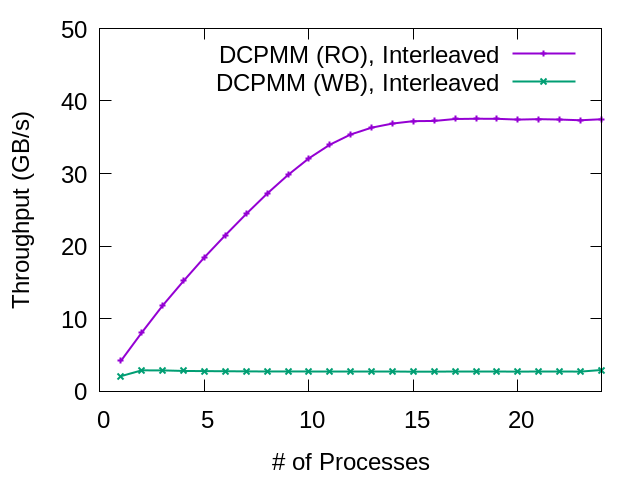}
		\caption{DCPMM (Interleaved)}
	\end{subfigure}
	\begin{subfigure}{0.33\linewidth}
		\includegraphics[width=\columnwidth]{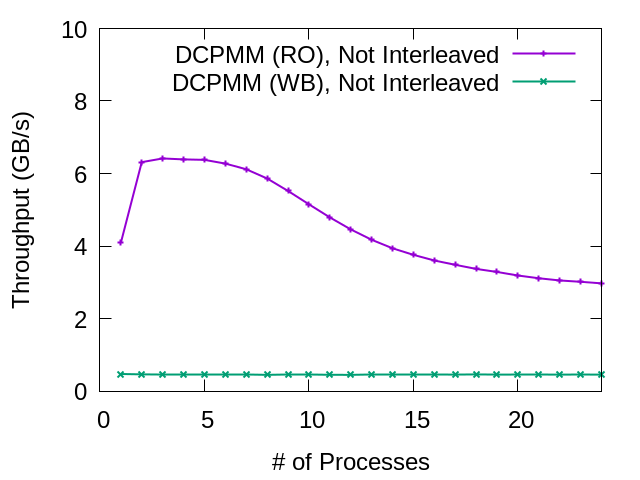}
		\caption{DCPMM (Non Interleaved)}
	\end{subfigure}
	\caption{The read/write memory bandwidths of DRAM and DCPMM. In the graphs, the results of the read latency are marked as RO (read-only), and those of the write latency are marked as WB (write-back).}
	\label{fig:bw}
	\end{center}
\end{figure*}

Figure \ref{fig:bw} shows the read/write bandwidths of DRAM and DCPMM, respectively.
As the number of the concurrent worker processes increased for read-only memory
access, the bandwidth of DRAM reached 101.3 GB/s at peak; on the other hand, the
bandwidth of DCPMM was 37.6 GB/s.
For memory access involving write-back,
the bandwidth of DRAM was 37.4 GB/s at peak, and that of DCPMM was 2.9 GB/s.
For read access, the throughput of DCPMM was 37.1\% of DRAM. For write access, it was 7.8\%.
The difference in read and write bandwidths is larger in DCPMM; it was approximately 13 times in DCPMM, while it was 2.7 times in DRAM.

With the interleaving of DCPMM disabled, the observed peak bandwidths were degraded to approximately 1/6 (i.e., 6.4 GB/s for read-only access, and 0.46 GB/s for write-back-involving access).
The number of the memory modules, being simultaneously accessed, was only one (i.e., 1/6 of the interleaved configuration).
Interestingly, as the number of concurrent worker processes increased,
the throughput of read access decreased by approximately 50\%.
A possible reason is that the internal buffering mechanism of DCPMM does not
work efficiently when the interleaving mechanism is disabled.
Its design is supposed to be optimized for interleaved memory accesses.

\subsection{Summary and Discussion}
Table \ref{tbl:summary} and Table \ref{tbl:summaryinter} summarize the key results of our experiments.
The advantage of DCPMM is the large capacity of a memory module (e.g., 128 GB,
256 GB and 512 GB), which is an order of magnitude greater than that of DRAM (i.e., typically up to 32 GB).
Its disadvantage is its modest read/write performance:
{\flushleft Latency:}
\begin{itemize}
	\item The read latency was approximately 374.1 ns, which was 400.1\% larger than that of DRAM.
	\item The memory access latency involving write back operations was approximately 391.2 ns, which was 407.1\% times larger than that of DRAM. Without interleaving, it was degraded to 458.4 ns.
\end{itemize}

{\flushleft Bandwidth:}
\begin{itemize}
	\item The read bandwidth of DCPMM was approximately 37.6 GB/s, which was 37.1\% of that of DRAM.
        \item The memory access bandwidth involving write back operations was approximately 2.9 GB/s, which was 7.8\% of that of DRAM.
\end{itemize}

The obtained performance numbers complement prior work. To make the
contribution of the paper clear within the page limit of the letter format, we
discussed comparison with prior work in the latter half of Section 1.

\begin{table}[t]
	\begin{center}
		\caption{The obtained performance numbers of interleaved DRAM and DCPMM}
		\label{tbl:summary}
		{\footnotesize
		\begin{tabular}{l l|r|r|r}
			\hline
			    & & \multicolumn{1}{c}{DRAM} & \multicolumn{1}{|c|}{DCPMM} & \multicolumn{1}{c}{Ratio} \\ \hline
			\multirow{2}{*}{Latency}              & Read-only  & 93.5 ns   & 374.1 ns   & 400.1\% \\
			                                      & Write-back & 96.1 ns   & 391.2 ns   & 407.1\% \\     \hline
			\multirow{2}{*}{Bandwidth}            & Read-only  & 101.3 GB/s &  37.6 GB/s &  37.1\% \\
			                                      & Write-back &  37.4 GB/s &   2.9 GB/s &   7.8\% \\
			\hline
		\end{tabular}
		}
	\end{center}
\end{table}

\begin{table}[t]
	\begin{center}
		\caption{The obtained performance numbers of interleaved and non-interleaved DCPMM}
		\label{tbl:summaryinter}
		{\footnotesize
		\begin{tabular}{l l|r|r|r}
			\hline
			    & & \multicolumn{1}{c}{Interleaved} & \multicolumn{1}{|c|}{Non-Interleaved} & \multicolumn{1}{c}{Ratio} \\ \hline
			\multirow{2}{*}{Latency}              & Read-only  & 374.1 ns   & 394.5 ns & 105.5\% \\
							      & Write-back & 391.2 ns   & 458.4 ns & 117.2\% \\     \hline
			\multirow{2}{*}{Bandwidth}            & Read-only  &  37.6 GB/s &  6.4 GB/s & 17.0\% \\
							      & Write-back &   2.9 GB/s &  0.46 GB/s & 15.9\% \\
			\hline
		\end{tabular}
		}
	\end{center}
\end{table}

\section{Conclusion}
In order to complement prior performance reports on Intel Optane DCPMM,
we conducted experiments using our own measurement tools.
We observed that the latency of random read-only access was approximately 374 ns.
That of random writeback-involving access was 391 ns. The bandwidths of
read-only and writeback-involving access for interleaved memory modules
were approximately 38 GB/s and 3 GB/s, respectively.

Many applications (e.g., especially large-scale HPC and AI workloads) will get benefit from a large capacity of main memory expanded by DCPMM.
However, a substantial performance gap between DCPMM and DRAM poses new challenges for system software studies.
We are currently conducting experiments using application programs and will report details in our future publication.

\section*{Acknowledgment}
We would like to acknowledge the support of Intel Corporation.
We also thank Dr. Jason Haga and other colleagues for their invaluable feedback.

\bibliographystyle{ieicetr}
\bibliography{ms}

\end{document}